\documentclass[twocolumn,showpacs,prl,amsmath,amssymb]{revtex4}
\usepackage{graphicx}
\usepackage{bm}
\begin{document}

\title{Cold Atom Optical Lattices as Quantum Analog Simulators for
Aperiodic One-Dimensional Localization Without Disorder}
\author{V.W. Scarola and S. \surname{Das Sarma}}
\affiliation{Condensed Matter Theory Center, 
Department of Physics, University of Maryland,
College Park, MD 20742-4111}

\begin{abstract}
Cold atom optical lattices allow for the study of quantum localization
and mobility edges 
in a disorder-free environment.  We predict the existence of an 
Anderson-like insulator with sharp mobility edges in a 
one-dimensional nearly-periodic 
optical lattice.  We show that the
mobility edge manifests itself as the early onset of pinning in center
of mass dipole oscillations in the presence of a magnetic trap which
should be observable in optical lattices.  
\end{abstract}
\maketitle

\newcommand\degrees[1]{\ensuremath{#1^\circ}}

Optical lattices incorporating ultracold atomic condensates are
rapidly becoming ideal quantum systems for studying various model
Hamiltonians developed earlier for studying solid-state 
phenomena.  This is primarily due to the extraordinary level
of precise tunability that experimentalists have achieved in
controlling the parameters (e.g. hopping, interaction and lattice
periodicity) of the optical lattice, which makes
it possible for the cold atom optical lattice to operate as an ideal
quantum analog simulator for various many-body condensed matter
Hamiltonians.  By contrast, ideal model Hamiltonians 
(e.g. Hubbard and Anderson models) often poorly describe
solid-state systems since experimental control over
complex condensed matter systems is, in general, quite limited.  In
addition solid-state systems are invariably contaminated by
unknown disorder, defects, and impurities whose effects are not easy to
incorporate in model Hamiltonians.  The cold atom optical lattices are
therefore becoming increasingly important in advancing our knowledge
about the quantum phase diagram and crossover in model many-body
Hamiltonians of intrinsic interest.  Examples include: the
Bose-Hubbard model \cite{Greiner}, the Tonks-Girardeau gas 
\cite{Paredes}, and the BEC-BCS
crossover \cite{cross}.  

In addition to studying strong correlation effects (e.g. the
superfluid-Mott insulator transition in the Bose-Hubbard model) in
many-body Hamiltonians, cold atom optical lattices
also offer ideal systems for studying quantum transport phenomena
including ballistic quantum transport \cite{Oskay,dArcy,Sadgrove} and quantum
localization \cite{Drese,Guidoni,Pezze,Huckans}.  The latter may be
more generally classified as metal-insulator transition 
phenomena with a direct relationship to the solid-state.  The distinction 
between a ``metal'' (i.e. a system with
finite resistivity at zero temperature) and an ``insulator'' (i.e. a
system with infinite zero temperature resistivity) is purely 
quantum.  Broadly speaking, there are four classes of
metal-insulator transitions in quantum lattice systems: Metal-band 
insulator transition in an ordered periodic lattice 
arising from the chemical potential
moving into energy band gaps; interaction induced metal-insulator
transition as in the Mott transition; disorder induced quantum
localization (i.e. Anderson localization \cite{Anderson}); and 
quantum localization in
aperiodic (but deterministic) potentials in disorder-free lattice
systems.  
 
In this paper, we establish that very
general aspects of the metal-insulator transition phenomena (in the
disorder-free environment) can be directly experimentally studied
in aperiodic cold atom optical lattices with the tuning of experimental
parameters leading to the observation of {\em both} band and
quantum (Anderson-like) localization in the same system but in different
parameter regimes.  Such an experimental study of localization or
insulating transitions in deterministic aperiodic systems is 
impossible in solid state lattice systems since disorder (which
leads to direct Anderson localization) is invariably present in solid
state systems overwhelming any subtle localization effects
arising from deterministic aperiodic potentials.  In particular,
all states are localized in one-dimensional systems in the 
presence of any disorder whereas one-dimensional aperiodic potentials
allow for the existence of extended quantum eigenstates.  This makes
one-dimensional optical lattice systems particularly interesting from
the perspective of localization studies in deterministic aperiodic
potentials since such studies in the corresponding one-dimensional
solid-state systems are essentially impossible due to disorder
effects.  We therefore consider aperiodic quantum localization 
in one-dimensional optical lattices, conclusively establishing the
feasibility of studying this unusual phenomenon in cold atom optical
lattices.      

The single-particle quantum localization problem in a deterministic 
quasiperiodic potential (i.e. two lattice potentials with mutually
incommensurate periods) has a long history \cite{Sokoloff,Last}.  In
particular, localization properties have been extensively studied in
the Harper (or, equivalently, Aubry) model which has an intriguing
self-dual point where the eigenstates form a multifractal Cantor set
spectra and are neither localized nor extended.  
Away from the dual point conventional wisdom dictates that all
states, as a function of the chemical potential, are either 
all extended or all localized, depending on the mutual strengths of the 
potential and hopping terms.  Such Harper model
type quasiperiodic potentials therefore do not allow for the existence
of a mobility edge separating extended states (above the mobility
edge) from localized states (below the mobility edge) which is the
hallmark of the Anderson localization transition in three-dimensional
disordered system.  Central to our work is the conclusive theoretical 
demonstration of a class of one-dimensional
optical lattice systems where the deterministic lattice potential
{\em does} allow for the existence of a mobility edge in one dimension
\cite{Azbel}, which cannot happen through Anderson localization with
disorder.  This class of models distinguishes itself from other models
discussed in the context of optical lattices \cite{Diener} 
through the formation of a metal-insulator mobility edge rather than a
metal-band edge.  We find that: 1) Direct numerical simulation and an  
analytic WKB approximation provide conclusive evidence for a 
rare metal-insulator mobility edge in a one-dimensional model, the nearly 
periodic Harper model. 2)  Transport measurements in suitably designed,  
one-dimensional optical lattices can exhibit the mobility edge. 

We consider spinless fermions (or equivalently hardcore bosons sufficiently 
near the Tonks-Girardeau regime) in the lowest band 
of a one-dimensional, tight binding 
lattice with external potentials:
\begin{eqnarray}
-u_{n+1}-u_{n-1}+(V_n+V_D f_n +\Omega n^2)u_n=Eu_n,
\label{H}
\end{eqnarray}
where the amplitudes $u_n$ multiply the Wannier states at sites $n$ in the
real space wavefunction $\Psi(x)=\sum_n u_n w(x-n)$.  
We work in units of the hopping matrix element, $t=1$, and 
lattice spacing of the primary lattice defining the tight 
binding problem, $a=1$, unless otherwise noted.  The statistics 
of spinless fermions
implicitly allow for an arbitrary on-site interaction in the above
single-band model.  In the absence
of an external potential the solutions form extended
states, $u_n=u_0\exp(in\phi)$, with band energies 
$E=2\cos(\phi)$, for $0\leq\phi\leq\pi$.  The band edges lie at
$E=\pm 2$.  In the presence of an 
oscillatory modulation of strength $V$, much weaker than the 
primary lattice, we can ignore
modifications to the hopping. In this limit we impose a secondary lattice:
$    
V_n=V\text{cos}(2\pi\alpha n)-V.
$
  For $\alpha$ irrational the additional potential establishes an
 incommensurate pseudorandom model, the 
Harper model (for $\Omega=0$ and $V_D=0$).  The potential $V_D f_n$
adds disorder where $f_n$ is a random number satisfying 
$0\leq f_n \leq 1$ for each site.  The confinement 
potential, $\Omega n^2$, applies to optical lattice systems.      

According to the Aubry-Andre
conjecture \cite{Aubry} the Harper model exhibits a 
metal-insulator transition at the self-dual point $V=2$.  For $V<2$
all states are extended while for $V>2$ all states localize (the
states at $V=2$ are critical).  The
localized states are characterized by a nonzero Lyapunov 
exponent (inverse localization length), $\gamma(E)$, where 
$u_n(E)\sim\exp(-\gamma n)$, and gaps in 
the energy spectra.  While exceptions to the 
Aubry-Andre conjecture have been rigorously proven for specific values of
$\alpha$ \cite{Avron}, we discuss here an additional and 
experimentally relevant counter example defined by:
$\alpha=m\pm\epsilon$, for integer $m$ and  
\begin{eqnarray}
N^{-1}\ll \epsilon \ll 1,
\label{lim}
\end{eqnarray}
with $N$ sites.  In the limit $N\rightarrow \infty$ the 
secondary lattice defines a
slowly varying, nearly-periodic potential.  A similar, slowly varying
potential has been considered in the context
of one-dimensional localization in quasiperiodic systems \cite{DasSarma}.   

\begin{figure}
\includegraphics[clip,width=3.0in]{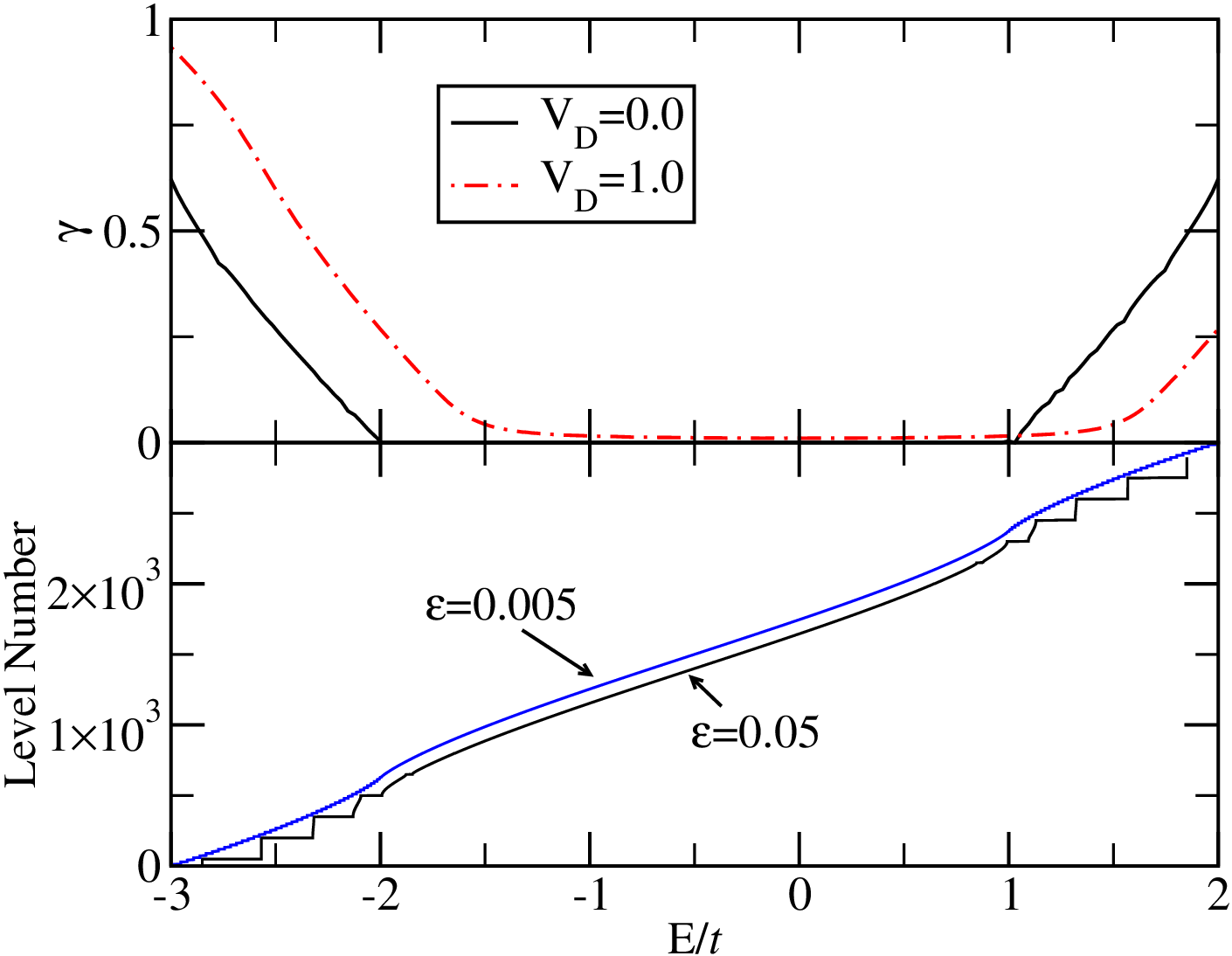}
\caption{Top panel:  Lyapunov exponent versus energy for 
the nearly-periodic Harper model, Eq.~(\ref{H}) with $V=0.5$, 
  $\epsilon=0.005$, and $\Omega=0$.  The solid (dot-dashed) line
  shows the disorder-free, $V_D=0$ (disordered, $V_D=1$) case.
  Bottom panel:  Level number versus energy of the disorder-free,
  nearly-periodic Harper model for two characteristic values of
  $\epsilon$ with $V=0.5$, $N=3000$, and $\Omega=0$.  The lower curve 
  is shifted downward by 100 
  levels for clarity.   
\label{gam}}
\end{figure}

In the limit defined by Eq.~(\ref{lim}) the eigenstates 
of Eq.~(\ref{H})  with $\Omega=0$ and $V_D=0$ display Anderson-like localization 
where we expect to find 
{\em only} extended states.  To see this consider $\gamma$ defined
in the limit, $N\rightarrow \infty$ \cite{Thouless}:
\begin{eqnarray}
\gamma(E_j)
=\frac{1}{N}\sum_n \ln \left\vert \frac{u_{n+1}}{u_{n}}\right\vert
=\frac{1}{(N-1)}\sum_{j\neq l}\ln\left\vert E_j-E_l\right\vert.
\label{defg}
\end{eqnarray}
The first equality allows us to use the transfer matrix method to 
calculate $\gamma$ for large system sizes.  The solid
line in the top panel of Fig.~\ref{gam} plots the Lyapunov exponent 
versus energy for $N=10^7$ and $V=0.5$.  
The additional potential, $V_n$, shifts 
the lower band edge to $E=-2-2V$ while leaving the upper band edge at
$E=2$.  We see extended states in the center of 
the band, $-2<E<2-2V$, with $\gamma=0$, as 
expected from the Aubry-Andre conjecture.
However, near the band edges, $-2-2V<E< -2$ and $2-2V<E<2$, the 
states localize, $\gamma>0$.  
The points $E=-2$ and $2-2V$ define mobility edges which are  
unexpected in one dimension but found in three-dimensional models with
disorder.  The localization is, in this sense, Anderson-like.  
We find that, for $N=10^7$, the mobility edges persist for rational {\em and}
irrational values of $\epsilon$ from $10^{-5}$ to $10^{-2}$.  We
conjecture that in the limit Eq.~(\ref{lim}) irrational numbers are
approximated by rational numbers up to a number much smaller than
$N^{-1}$.  For $N\rightarrow \infty$, the spectra can contain an
infinite number of infinitely small gaps and therefore localized
states eliminating the distinction between an incommensurate and
commensurate system \cite{Avron,Sokoloff}. 

The unexpected insulating behavior coincides with a devil's staircase-like
structure in part of the energy spectrum \cite{Azbel,Sokoloff}.  The 
second equality in Eq.~(\ref{defg}) shows that a degeneracy at $E_j$ 
supports non-zero $\gamma(E_j)$.  The lower panel of 
Fig.~\ref{gam} plots the 
level number as a function of energy as determined by exact
diagonalization of Eq.~(\ref{H}) for $N=3000$, $V=0.5$, $V_D=0$, and 
$\Omega=0$.  The top line
($\epsilon=0.005$) shows smaller gaps and narrower steps than the bottom line
($\epsilon=0.05$) suggesting that the localized states develop a gapless      
insulator in the limits $\epsilon\rightarrow 0$ and $\epsilon \gg N^{-1}$.

We can understand the insulating states in a ``semi-classical''
approximation where $\epsilon$ plays the role of $\hbar$.  We analyze 
the behavior of each regime as a function of energy.  At low energies,
$E<-2$, the 
slowly varying potential confines low
energy states near the potential minima defined by $V_n$.  Very 
little tunneling between minima forces localization.  
Intermediate energies,$-2<E<2-2V$, see a smaller 
barrier between minima allowing for 
extended states and, therefore, the first mobility edge 
at $E=-2$.  A second mobility edge forms at   
$E=2-2V$ when states localize at the 
secondary lattice maxima.  
At first this seems counterintuitive but can be understood in a WKB
approximation based on the slowly varying nature of $V_n$.  A similar
analysis was performed for a different model in Ref.~\cite{DasSarma}.  Our results 
show that the high energy
states, $E>2-2V$, moving energetically above the lattice slow 
when passing secondary lattice maxima to force localization.  We have checked
that our analysis based on the WKB approximation reproduces the
solid line in the upper panel of Fig.~\ref{gam}.  As an additional
check we can, in a continuum approximation \cite{Sokoloff,Wimberger}, 
define a position variable, $n\rightarrow \tilde{x}$, and a 
difference operator, $u_{n+1}+u_{n-1}\rightarrow 2\text{cos}(p)u(\tilde{x})$
(with $p\equiv i\partial/\partial \tilde{x}$), to give the 
semi-classical Hamiltonian: $H_{\text{CL}}= -2\text{cos}(p)+V(\tilde{x})$,
with the replacement $V_n\rightarrow V(\tilde{x})$.  The phase trajectories of
$H_{\text{CL}}$ produce extended and localized states (and therefore
mobility edges) in the regimes obtained in Fig.~\ref{gam}. 

We now discuss the possibility of observing this unique type of
localization.  In the solid state a necessary correction to
the Harper model includes disorder where we add to $V_n$ a potential of the form:
$V_{\text D}f_n$.  For $V=0$ (and $\Omega=0$) this defines the one-dimensional Anderson
model where we expect all states to localize for arbitrary $V_D$.
However, for $V\neq 0$, the states (otherwise extended in the $V_{\text D}=0$
case) have a 
small localization length which could allow some 
remnant of a mobility edge. The
dot-dashed line in the upper panel of Fig.~\ref{gam} plots $\gamma$
for the same parameters as the solid line but 
with $V_{\text D}=1.0$.  We find that a finite amount of
disorder obscures the position of the remnant-mobility edges while
localizing all states.      

In what follows we consider an essentially disorder-free manifestation of
Eq.~(\ref{H}): one-dimensional, cold atom optical lattices.     
The interference of appropriately detuned lasers of wavelength
$\lambda=2a$ can give rise to our tight binding lattice with
a sufficiently strong lattice height $V_L$. To create a 
secondary modulating potential, $V_n$,
consider an additional pair of lasers at angles $\theta$ and
$\pi-\theta$ to the primary lattice with wavelength $\lambda '$ and 
amplitude $V_L '$.  
The additional lasers interfere
to modulate the energy of the $n$th site by: 
$V_L '\text{cos}(2\pi n\alpha )
\int_{-\infty}^{\infty}\vert w(u) \vert^2
\text{cos}(2\pi u\alpha )du-V_L '$,
where  $\alpha=(\lambda/\lambda ')\text{cos}(\theta)$.  For small
angles we can retrieve, up to an overall constant, our nearly-periodic
Harper model with $m=\lambda/\lambda '$ an integer and 
$\epsilon\approx-\lambda\theta^2/2\lambda '$.  For realistic parameters:
$V_L=5 E_R$, $V_L '=0.1E_R,\theta=\degrees{5}$, and 
$\lambda=\lambda'$ (where $E_R$ is the 
photon recoil energy), we find $t\approx 0.065 E_R$, $V\approx
0.055 E_R$, and $\epsilon\approx 0.004$ yielding the appropriate parameter 
regime.  Furthermore, we find that, in the limit of Eq.~(\ref{lim}),   
fluctuations in the relative phase do not alter the position of the mobility
edge.  We now include an 
important modification to the model which accounts for realistic
finite size effects.

\begin{figure}
\includegraphics[clip,width=3.0in]{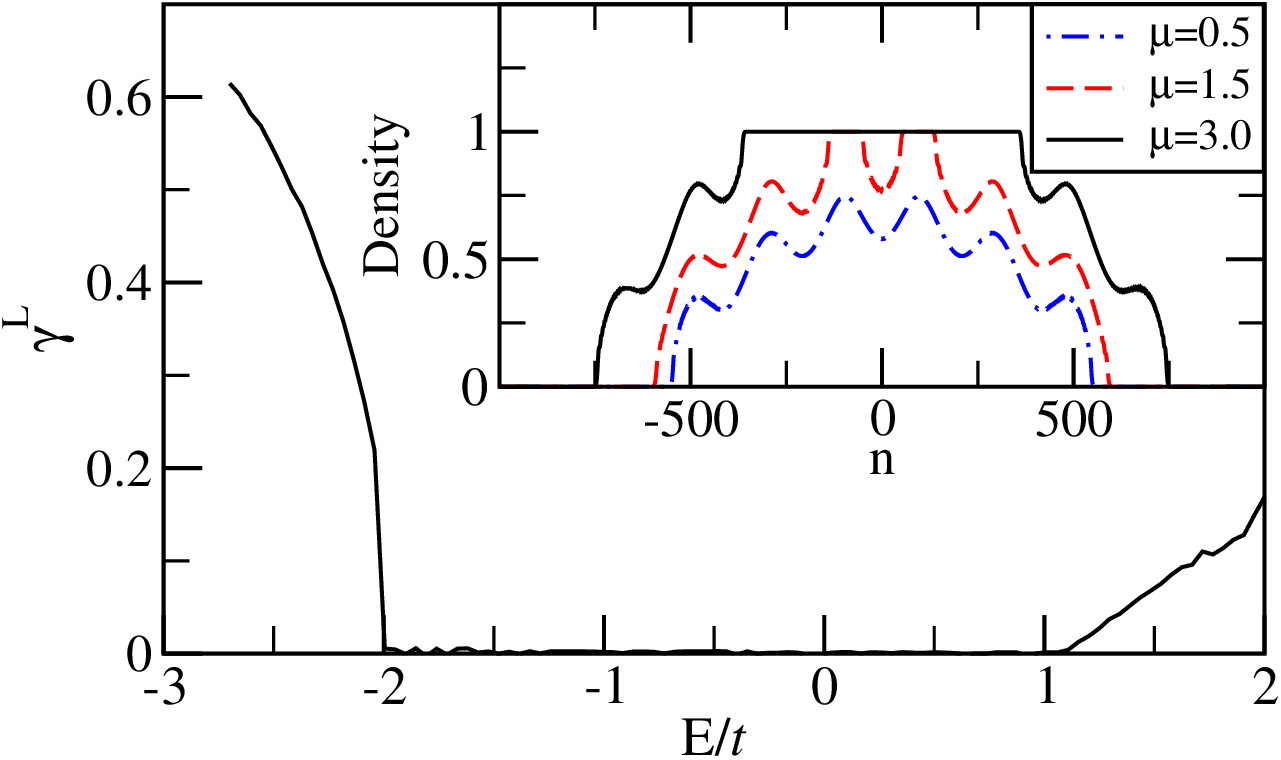}
\caption{The local Lyapunov exponent versus energy for the same
  parameters as the upper panel of Fig.~\ref{gam} but with an additional
  parabolic confinement $\Omega=10^{-5}$ and no disorder, $V_D=0$.
  The inset shows the
  normalized density of the same system as a function of 
  lattice number for several different chemical potentials.  
\label{parab}}
\end{figure}

A crucial addition to the Harper model in optical lattices 
is the parabolic confinement: $\Omega n^2$, which leads to a finite 
particle number.  We find
that weak confinement leaves the mobility edges intact.  To see this consider
the local Lyapunov exponent:      
$
\gamma^L(E_j)=(2N_{\text{CL}}+1)^{-1}
\sum_{n= -N_{\text{CL}}}^{N_{\text{CL}}} 
\ln\vert u_{n+1}/u_n\vert,
$
where the semiclassical limits of the parabolic
trap define the number of states participating in 
transport, $2N_{\text{CL}}+1$.  The classical turning points give 
$N_{\text{CL}}=2\vert x_{\text{CL}}(E) \vert$ and Eq.~(\ref{lim}) 
becomes: $(2N_{\text{CL}})^{-1}\ll \epsilon \ll 1$.  To determine 
$x_{\text{CL}}$ we set $p=0$ in $H_{\text{CL}}$ with
$V(\tilde{x})=V\text{cos}(2\pi \alpha \tilde{x} )-V+\Omega \tilde{x}^2$.  
For $\Omega\sim 10^{-5}$ we find $2N_{\text{CL}}\sim 10^3$.  In the
limit $\Omega\rightarrow 0$ we retrieve the usual Lyapunov exponent, 
$\gamma^L \rightarrow \gamma$.
Fig.~\ref{parab} plots the local Lyapunov exponent as a function of
energy for $N=10^7$, $\epsilon=0.005$, $V=0.5$, $V_D=0$, and 
$\Omega =10^{-5}$.  The 
mobility edges remain even with a reduced number of states comprising 
the system.    
The inset shows the normalized density profile as a function of site
number for three different chemical potentials, $\mu$.  At zero
temperature we include states with $E\leq\mu$.  For $\mu=0.5$
(dashed-dotted line) we find extended states with some modulation due
to $V_n$.  For
$\mu=1.5$ (dashed line) we have crossed the mobility edge and the
density pins to unity at some lattice sites.  Here the formation
of a mesoscopic version of the Anderson-like gapless insulator 
fixes the density.  For $\mu=3.0$ we enter the band
insulator regime which fixes a large fraction of the 
states at integer density.

Dipole oscillations in harmonically confined atomic gases serve as a 
direct probe of localization \cite{Pezze,Huckans}.  A small shift in 
the center of mass results in harmonic oscillations 
in the absence of an external lattice.
The presence of one or more weak lattices allows for weakly localized states 
which can suppress oscillations and lead to an effective under-damping of
the center of mass motion.  The addition of strongly localized states   
can, in the absence of dissipation, eventually pin the center of mass 
to effectively over-damp the center of mass oscillations.  Strong experimental
and theoretical evidence supports the possibility that band
localization has indeed been observed in fermionic, one-dimensional
optical lattices \cite{Pezze}.  Similar evidence also suggests such behavior
for strongly interacting bosons \cite{Huckans}.    

\begin{figure}
\includegraphics[clip,width=3.0in]{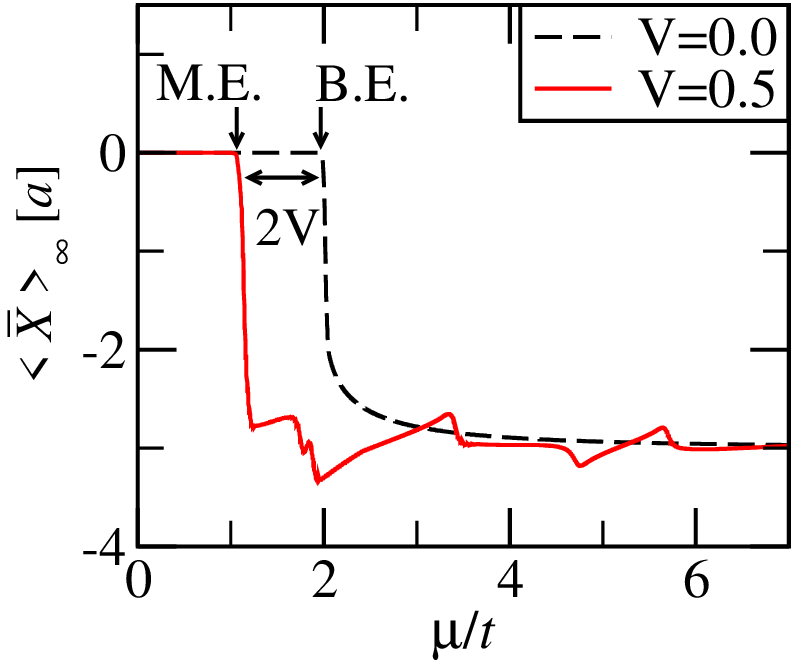}
\caption{  The long time average of the center of mass position 
  as a function of chemical potential after an initial 
  displacement of three sites, $\Delta=-3$, with no disorder, $V_D=0$.  The dashed 
  line is calculated from the bare tight
  binding model with no secondary lattice, $V=0$, 
  and $\Omega=10^{-5}$.  B.E labels the
  upper band edge.  The solid line is calculated for the same
  parameters but with $V=0.5$.  The early onset of the gapless insulator occurs at
  the upper mobility edge, labeled M.E., while the upper band 
  edge remains at $\mu=2$.
\label{avx}}
\end{figure}

We now study the onset of
the gapless Anderson-like insulator and its effect on center 
of mass oscillations.
Consider the center of mass to be displaced $\Delta$ lattice sites 
at some initial time $T=0$.  For extended states, the center of mass 
position, $\bar{X}(T)$, averages to zero for long times  
while localized states should pin the center of mass position,  
$\bar{X}\sim \Delta$.  The center of mass position can, for some
parameters, demonstrate complex, damping-like 
behavior as function of
time making a damping constant ill-defined.  To extract a simple 
quantity to be compared with experiment 
we calculate the long time average of the
center of mass position, $<\bar{X}>_{\infty}$, as a function of chemical
potential by diagonalizing Eq.~(\ref{H}) with a parabolic
potential, $\Omega=10^{-5}$, for $N=3000$, $\Delta= -3$,
and $V_D=0$.  As an intermediate step we require degenerate 
eigenstates (localized
at the edges) to simultaneously diagonalize the parity operator since our 
system possess reflection symmetry about the origin.  
The dashed line in 
Fig.~\ref{avx} plots $<\bar{X}>_{\infty}$ as a function of chemical
potential in the absence of a secondary lattice, $V=0$.  For $\mu<2$ the 
extended states perform several oscillations about the trap center but
over long times average to zero displacement.     
Above the band edge (labeled B.E.), for $\mu>2$, localized states 
near the edge pin the center of mass near $\Delta$.  For $\mu \gtrsim 3$
the system never leaves its initial position.  

A second weaker
lattice causes a mobility edge to form 
energetically below the band edge.  The solid
line in Fig.~\ref{avx} plots the same as the dashed line but with a
second lattice, $V_n$, with $V=0.5$ for chemical potentials near the upper
mobility edge (labeled M.E.).  $<\bar{X}>_{\infty}$ remains 
zero where we expect
extended states but pins near $\Delta$ for $\mu>2-2V$.  The
mesoscopic version of the gapless insulator results in the early onset 
of pinning in the regime $2-2V<\mu<2$ and Eq.~\ref{lim}.  Furthermore,
the localized
states with the additional lattice, $V=0.5$, also display weak
periodicity in $<\bar{X}>_{\infty}$ as a function of $\mu$.  These 
oscillations correspond to the
chemical potential passing through peaks and valleys in the
corrugated confinement potential. 

Fluctuations in the lattice depth can soften the otherwise sharp
mobility edge.  The quantity of
interest, $2V/t$, can fluctuate wildly with only moderate changes in $V_L$ 
at extremely large lattice depths.  To see this consider 
an approximate expression in terms of the hopping 
extracted from an analysis of the related Mathieu problem:
$V/t\approx (\sqrt{\pi}V/4)(V_L/E_R)^{-3/4}\exp(2\sqrt{V_L/E_R})$.  A relative 
error in $V$ and $V_L$, $R_V$ and $R_{V_L}$ respectively,
propagates to a relative error in $2V/t$: 
$[R_V^2+R_{V_L}^2(3/4-\sqrt{V_L/E_R})^2]^{1/2}$.  
We have checked that this formula is
quantitatively accurate for $V_L\gtrsim 5 E_R$ by comparing with error
derived numerically from the exact tunnelling.  We find that for
$R_V=R_{V_L}=5 \%$ the relative error in $2V/t$ 
remains below $20 \%$ for $V_L<20E_R$.  
  
We note that additional time dependence in the model discussed 
here possesses other applications.  We take $H_{\text{CL}}$ as a good 
approximation to the nearly-periodic Harper model in the limit Eq.~(\ref{lim}).   
In the presence of a pulsed secondary lattice:   
$V\propto \sum_{j}\delta\left(T-jT_0\right)$, where for integer $j$
the secondary lattice oscillates with period $T_0$, we simulate 
the kicked Harper model via $H_{\text{CL}}$.  The
kicked Harper model exhibits chaotic behavior with the ``classical''
to quantum crossover controlled by $\epsilon$.
       
We have explicitly demonstrated the existence of a mobility edge
(and the associated, unusual metal-insulator transition in a
deterministic disorder-free environment) in suitably designed
aperiodic cold atom optical lattice systems.  The deterministic 
aperiodic background potential in these optical lattices leads to
exotic and nontrivial energy eigenstates 
dependent on the relationship between irrational numbers and their 
rational approximations.  The ensuing quantum localization occurs in
the absence of disorder and therefore 
distinguishes itself from Anderson localization which, in the solid
state, masks the presence of mobility edges formed from quasiperiodic
potentials in one dimension.

We thank K. Park and G. Pupillo for valuable
discussions.  This work is supported by NSA-LPS and ARO-ARDA.



\end{document}